\documentclass{article}

\usepackage{arxiv}

\usepackage[utf8]{inputenc} 
\usepackage[T1]{fontenc}    
\usepackage{hyperref}       
\usepackage{url}            
\usepackage{booktabs}       
\usepackage{amsfonts}       
\usepackage{nicefrac}       
\usepackage{microtype}      
\usepackage{lipsum}
\usepackage[numbers]{natbib}

\usepackage{color}
\providecommand{\s}[1]{{\color{black}#1}}
\usepackage{textcomp}
\newcommand{\textapprox}{\raisebox{0.5ex}{\texttildelow}}
\newcommand*\samethanks[1][\value{footnote}]{\footnotemark[#1]}
\usepackage{graphicx}

\title{Hierarchical, rotation-equivariant neural networks to select structural models of protein complexes}

\author{
  Stephan Eismann\thanks{Equal contribution} \\
  Department of Applied Physics\\
  Stanford University\\
  \texttt{seismann@stanford.edu} \\
   \And
 Raphael J.L. Townshend\samethanks \\
  Department of Computer Science\\
  Stanford University\\
  \texttt{raphael@cs.stanford.edu} \\
     \And
 Nathaniel Thomas\samethanks \\
  Department of Physics\\
  Stanford University\\
  \texttt{nthomas103@gmail.com} \\
       \And
 Milind Jagota \\
  Department of Electrical Engineering\\
  Stanford University\\
  \texttt{mjagota@stanford.edu} \\
       \And
 Bowen Jing \\
  Department of Computer Science\\
  Stanford University\\
  \texttt{bjing@stanford.edu} \\
         \And
 Ron O. Dror \\
  Department of Computer Science\\
  Stanford University\\
  \texttt{rondror@cs.stanford.edu} \\
}

\begin{document}
\maketitle

\begin{abstract}
Predicting the structure of multi-protein complexes is a grand challenge in biochemistry, with major implications for basic science and drug discovery. Computational structure prediction methods generally leverage pre-defined structural features to distinguish accurate structural models from less accurate ones. This raises the question of whether it is possible to learn characteristics of accurate models directly from atomic coordinates of protein complexes, with no prior assumptions. Here we introduce a machine learning method that learns directly from the 3D positions of all atoms to identify accurate models of protein complexes, without using any pre-computed physics-inspired or statistical terms. Our neural network architecture combines multiple ingredients that together enable end-to-end learning from molecular structures containing tens of thousands of atoms: a point-based representation of atoms, equivariance with respect to rotation and translation, local convolutions, and hierarchical subsampling operations. When used in combination with previously developed scoring functions, our network substantially improves the identification of accurate structural models among a large set of possible models. Our network can also be used to predict the accuracy of a given structural model in absolute terms. The architecture we present is readily applicable to other tasks involving learning on 3D structures of large atomic systems.
\end{abstract}

\section{Introduction}
Proteins bind to one another in specific ways to form protein complexes. The resulting complexes are essential for virtually all cellular processes, and the targeted blocking of protein-protein interactions is a key strategy in modern drug design \citep{Esmaielbeiki2016a}. Determining structures of protein complexes experimentally is often difficult and time-consuming, placing a premium on computational structure prediction \citep{Vakser2014}. Unfortunately, computational structure prediction for protein complexes has also proven difficult—much more so than for individual proteins.  

The process of predicting the structure of a protein complex given structures of the individual proteins involved, known as protein docking, generally involves two steps. First, the configurational space of the interacting proteins is sampled to produce a list of candidate models (hypothetical 3D structures of the complex). Second, each candidate model is assigned a score using a scoring function intended to assign the best scores to the most accurate models (i.e., those that would most closely match the experimentally determined structure, were it available). Despite the availability of many software packages designed for this purpose \citep{Chen2003, Venkatraman2009a, DeVries2010, Torchala2013, Szilagyi2014, Schindler2017, Kozakov2017,Quignot2018, Marze2018}, identifying accurate models out of a large set of candidate models has proven to be a major bottleneck for successful protein docking. 

Existing scoring functions generally leverage hand-designed, local features to assess the accuracy of a model. Widely used scoring functions can be classified as physics-inspired or statistical. Physics-inspired scoring functions include terms accounting for interatomic interactions and desolvation energy \citep{Vakser1994, Mandell2001, Pierce2007}. Statistical scoring functions are based on a probability distribution of defined spatial features (e.g. interatomic distances and bond orientations) computed for a set of known protein complex structures \citep{Kozakov2006, Mintseris2007, Dong2013, Andreani2013}. Recently, several scoring functions have been developed using machine learning methods \citep{Geng2019, Wang2019, Cao2020}. These also generally leverage pre-computed features as inputs. The use of pre-computed features can allow algorithms to directly focus on relevant information for predicting model accuracy. At the same time, all of the above methods might not consider certain features or characteristics of the atomic structure that are in fact relevant to identifying accurate models but have never been recognized as such.
In this paper, we explore whether we can improve solutions to the scoring challenge by considering all atoms of a model, without the use of hand-designed features. 

We present a neural network that learns directly from the 3D positions of all atoms to distinguish accurate models from less accurate ones, without making any assumptions about what characterizes a favorable protein–protein interface or what features are relevant to identifying an accurate model of a protein complex. The key challenge of designing such a machine learning method is to prevent an explosion of parameters to be fit, because the amount of data available for training (i.e., to fit these parameters) is limited by the number of experimentally determined structures. We also wish to represent the atom positions in a structural model as precisely as possible.

We address these challenges by designing a neural network architecture specifically for machine learning tasks involving large molecular systems. Our network architecture is based on convolutional filters that are precisely equivariant to rotation, translation, and permutation of the atomic coordinates that serve as inputs to the network. These physical symmetries are conserved over multiple hierarchical layers that we use to learn features at different levels of structural coarseness and to aggregate information globally. The equivariance property of our network significantly reduces the number of model parameters with no loss of model expressiveness. It also eliminates the need for rotational data augmentation and ensures that the orientation with which structural models are presented to the network does not matter.

This hierarchical, equivariant network architecture enables us to learn directly from a complete atomic representation of the protein complex, without making any assumptions about which features are or are not relevant to identifying accurate structural models. We demonstrate that a trained network, which we term PAUL, can pick out accurate models among a large set of candidate models for a given protein complex. When combined with existing scoring functions, PAUL substantially improves ranking performance and enriches the number of accurate models among those selected. We further demonstrate that a network can be trained to perform quality assessment---that is, to predict the accuracy of a given protein complex model in absolute terms. A webserver that implements PAUL is available at \url{drorlab.stanford.edu/paul.html}.

\section{Methods}
\subsection{End-to-end learning on structures of protein complexes}
\label{sec:lrmsd}
We present a method that allows effective end-to-end learning from large atomic structures. Atomic structure here means the most direct representation of a molecular system's 3D geometry: a list of the 3D coordinates and the element type of each atom. End‐to‐end refers to the concept that we do not manually define, select, or precompute any features that we believe are relevant.     

\begin{figure}[ht]
\begin{center}
\includegraphics[width=0.7\linewidth]{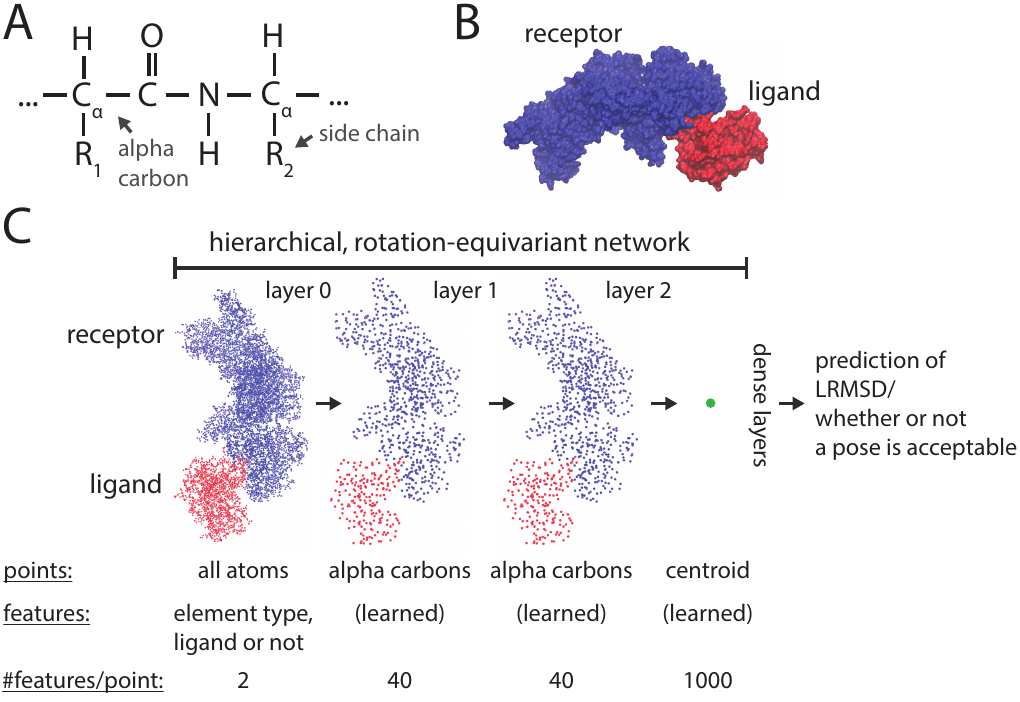}
\caption{\textbf{Learning from entire protein complex structures without prior assumptions.} 
(\textbf{A}) Proteins consist of chains of amino acid residues. Each residue has a central alpha carbon $C_\alpha$.  (\textbf{B}) Two proteins forming a protein complex. (\textbf{C}) Given 3D coordinates and element types of every atom as initial input, the network learns to predict either the ligand root mean square deviation (LRMSD) of a structural model or whether or not the model is acceptable (i.e., sufficiently accurate). Internally, each of the first three network layers learns a specified number of features for each of a specified set of points. Starting with all atoms, we use alpha carbons as the intermediate set of points before we output one large feature vector for the entire complex. This feature vector is a learned fingerprint of the atomic structure and serves as the input to four standard, fully-connected neural network layers, the last of which has a single scalar output.}
\label{fig:schematic}
\end{center}
\vskip -0.1in
\end{figure}

Figure~\ref{fig:schematic} illustrates the architecture of our deep learning network. The goal is to predict the accuracy of a given hypothesized protein complex model. We measure this accuracy in terms of the ligand root mean square deviation (LRMSD), a single scalar. The LRMSD of a model is calculated based on the atomic coordinates of the ligand (one specified protein among the two proteins) after the receptor (the second protein) has been aligned with the experimental structure. Figure~\ref{fig:schematic}B shows a protein complex in which receptor and ligand are highlighted. We use LRMSD because it is widely used in the structural biology community to measure the accuracy of a  protein complex model \citep{Lensink2007}. We train networks to perform two different tasks. One, which we term the regression task, is direct prediction of a model's LRMSD. The other, which we term the classification task, is prediction of whether or not a model is ``acceptable.'' We define an acceptable model as one with $\mathrm{LRMSD}<10$ \AA, in line with the assessment criteria of the \s{CAPRI protein docking experiment} \citep{Lensink2007}. We note that our neural network architecture can be trivially adapted to output a vector of any desired length---for example, to predict multiple properties at the same time. 

The structural model of a protein complex is defined by a list of atoms. Each atom's entry contains the atom's 3D coordinates, its chemical element, and a Boolean flag that indicates whether the atom is part of the receptor protein or part of the ligand protein. We represent carbon, oxygen, nitrogen, sulfur, and hydrogen atoms. The chemical element is specified by a vector of length 5 in which one entry is 1 and all others are 0 (one-hot encoding). We include hydrogen atoms as they play a vital role in the physics governing protein-protein interactions. The Boolean flag allows the network to recognize the binding interface between the two proteins. 

To enable end-to-end learning on these large molecules, we combine (i) neural networks equivariant to rotation, translation, and permutation, (ii) a novel, hierarchical learning approach, and (iii) nearest-neighbor convolutions. These three components, which we describe in the following sections, specifically cater to learning on atomic systems.  

\subsubsection{Equivariant neural networks}
Three-dimensional convolutional neural networks (3D CNNs) have been applied to several problems in structural biology, including the prediction of protein structure, protein-protein interfaces, and protein-ligand binding  \citep{Wallach2015, Ragoza2017, Torng2017,Jimenez2017, Derevyanko2018, Pages2019, Townshend2019}. Although powerful, conventional 3D CNNs are suboptimal in two ways when learning to evaluate the accuracy of large atomic structural models. First, conventional CNNs operate on a discretized grid, such that representing atomic coordinates with very high precision requires a very fine grid and thus a very large set of inputs to the network. Second, the convolutional filters of conventional 3D CNNs are not rotationally equivariant, posing a challenge to learning properties that are known to be invariant under rotations in 3D space, such as the accuracy of a structural model. In certain cases, one can solve this latter problem through a canonical alignment of the input data, but complexes of different proteins do not lend themselves naturally to such an alignment. One can also rely on rotational data augmentation---that is, rotating each training example at many different random angles---but this provides an imperfect approximation to rotation invariance, complicates the learning task, and poses challenges to capturing structural features involving finer-scale features that might appear at different orientations relative to one another in different structures (sometimes described as `patchwise' symmetry).

To account for the fact that the laws of physics that govern inter- and intramolecular interactions are invariant to translations and rotations, \citet{Behler2007} developed basic radial and angular symmetry functions in order to model the potential energy of small molecules through neural networks. These were further extended in SchNet \citep{Schutt2017} and ANI-1 \citep{Smith2017}. A number of recent publications have proposed neural network architectures that account for translational and rotational symmetries based on tools from group representation theory \citep{Cohen2016, Thomas2018, Kondor2018, Anderson2019}. Specifically, tensor field networks \citep{Thomas2018} have two key properties that distinguish them from conventional CNNs: they operate on a set of points with coordinates in 3D space (not a discretized grid), and they are automatically equivariant to rotations, translations, and permutations of those points.

Informally, a function (such as a neural network layer) is equivariant to some transformation (such as a rotation or translation) if applying the transformation to the input is the same as applying this transformation to the output. The classic example of this is a standard convolutional neural network:  a translation of the input to a convolutional layer by an integer number of pixels is equivalent to the same translation of the output. Equivariant networks enjoy the property that when each layer is equivariant, the whole deep network is automatically equivariant.  Equivariant networks eliminate the need for data augmentation with respect to the corresponding symmetry, and furthermore ensure local or `patchwise' symmetry at every part of the space (not only a single global symmetry).

Formally, a function $\mathcal{L} : \mathcal{X} \rightarrow \mathcal{Y}$ is {\em equivariant} with respect to a group $G$ and its representations $\mathcal{D}^\mathcal{X}$ and $\mathcal{D}^\mathcal{Y}$ if 
\begin{equation}
\mathcal{L} \circ \mathcal{D}^\mathcal{X}(g) = \mathcal{D}^\mathcal{Y}(g) \circ \mathcal{L}
\end{equation}
holds for all $g \in G$.  Equivariant neural networks bake in the prior that a symmetry with respect to $G$ always holds.  We expect the guarantee of equivariance to be a more powerful property the larger $G$ is (where the sense of `larger' in the cases we care about means the dimension of a continuous group $G$ when it is considered as a manifold ---for example, the group of 2D rotations is dimension 1, and the group of 3D rotations is dimension 3).

Concretely, we can represent inputs as $V_{am}$ (the coordinates and other features associated with each point), where $a$ indexes the points and $m$ is an index over the group representation (and in general will be broken up into multiple indices based on the internal structure of the representation, as described in \citet{Thomas2018}). This means that $L$ is equivariant if 

\begin{equation}
\sum_{a'm'}L_{am}^{a'm'}\left(\sum_{m''}D^\mathcal{X}_{m'm''}(g)V_{a'm''}\right) 
= \sum_{a'm'}D^\mathcal{Y}_{mm'}(g)\sum_{m''}L_{am'}^{a'm''}\left(V_{a'm''}\right), 
\end{equation}

for all $g \in G$, where $D$ denotes the corresponding matrix representations of $G$.

\subsubsection{Three-dimensional rotation- and translation-equivariant layers}

Mathematical  definitions of the tensor field network layers that we use here and proofs of their equivariance can be found in \citet{Thomas2018}.

To implement an equivariant convolution with respect to the group of proper 3D translations and rotations (the special Euclidean group $SE(3)$), we decompose convolution filters into a truncated series of spherical harmonics $Y$ with learnable radial functions $R$, which is combined with the input using a tensor product operation:
\begin{equation}
L_{a,c,(l,m)}^{a,c,(l',m')}\left(V_{a'',c,(l'',m'')}\right) 
= C^{(l', m')(l'', m'')}_{(l, m)} R_{c}^{(l')}(r_{aa''}) Y^{(l',m')}(\hat{r}_{aa''}) V_{a'',c,(l'',m'')},
\label{conv_defn}
\end{equation}
where $\vec{r}_{aa'}$ is the vector between points $a$ and $a'$, $C$ are the Clebsch-Gordan coefficients for the group of 3D rotations $SO(3)$ (implementing the tensor product), $(l, m)$ are the indices corresponding to the representation of $SO(3)$, and $c$ is the feature index.  Truncation of this series in $l$ is necessary because normally the series is infinite.

Additionally, we use self-interaction layers (analogous to 1x1 convolutions in CNNs) and nonlinear scaling of magnitudes (where we apply a nonlinear function to the norm over the representation index).  We also include a simple normalization layer, which includes a sum over the representation index in addition to a sum over the filter index (as in a typical normalization layer):
\begin{equation}
\frac{V_{a,c,(l,m)}}{\sqrt{\sum_{cm}|V_{a,c,(l,m)}|^2}}
\end{equation}
This normalization was found to aid in training convergence and is equivariant, as the representations of $SO(3)$ that we use are unitary.

\subsubsection{Hierarchical aggregation of information} \label{sec:hierarchical}
Rotation-equivariant convolutions do not require the set of input points to be the same as the set of output points \citep{Thomas2018, Kondor2018}; that is, $a$ and $a''$ do not have to range over the same sets in Eq.~\ref{conv_defn}. We take advantage of this fact to output at only a subset of points, a subsampling operation analogous to the stride in standard CNNs. We expect this to be useful for efficiently recognizing patterns at different scales of the protein structure.

Note that if we used such subsampling with rotation-{\em invariant} convolutions (such as those in \citet{Schutt2017}), this would eliminate lower-level orientation information, making it inaccessible to later network layers.  We call the largest $l$-value in the truncated spherical harmonic series the {\em maximum rotation order}, and it determines the angular resolution of the geometric information that is passed to later layers.

In the case of proteins, we start by considering all atoms and aggregate the information at the level of alpha carbons, where the alpha carbon is a canonical central atom in each amino acid (Fig.~\ref{fig:schematic}A). This reduces the number of points by a factor of \textapprox20 compared to the full set of atoms. We finally output a single vector assigned to the centroid of all atoms. This feature vector is the input to four standard, fully-connected layers with a single scalar output (Fig.~\ref{fig:schematic}C).

\subsubsection{Nearest-neighbor convolutions}
Naive use of the convolution layer calculates all pairwise interactions between points and therefore has $\mathcal{O}(N^2)$ complexity in the number of points $N$. However, the laws of physics are local; that is, the effects of physically relevant forces diminish with distance. As such, we reduce the complexity to $\mathcal{O}(N)$ by only calculating interactions for a fixed number $K$ of nearest neighbors of each point.  $K$ here is analogous to the area or volume of the convolution window for standard discrete 2D or 3D CNNs.  The overall computational complexity of our model thus scales linearly with the number of atoms. \s{We examine the influence of varying K in Figure S7. In all other experiments, we choose a value of $K=40$.} 

\subsection{Dataset} \label{sec:dataset}
\s{For training, we use protein complexes for which the Protein Data Bank \citep{Berman2000} includes experimentally resolved structures of the two proteins in complex as well as structures of the individual proteins. We generate models of protein complexes using the docking software ATTRACT \citep{Schindler2017} based on the experimentally determined structures of the individual proteins in their unbound state. We use complexes from docking benchmark 4 (DB4)~\citep{Hwang2010} for training (142 complexes) and validation (21 complexes). DB4 and its successor docking benchmark 5 (DB5)~\citep{Vreven2015} are widely used datasets specifically curated to benchmark protein docking methods as well as scoring functions used for protein docking.

We test our trained networks on two datasets. The first dataset contains 50 complexes that are part of DB5 but not DB4. Following the training procedure, we generate models of protein complexes with ATTRACT using the unbound structures of the individual proteins. The second dataset consists of 98 protein complexes from PPI4DOCK~\citep{Yu2016a}. PPI4DOCK provides experimentally resolved complex structures and homology models of the individual proteins in their unbound state. For this second dataset, we use the homology models of the individual proteins to generate models of protein complexes with ATTRACT. 

Our general sampling procedure considers rigid protein backbones. For testing, we also consider models generated with backbone flexibility. We include additional details regarding the datasets and sampling procedure in Supporting Information.
}

\subsection{Training} \label{sec:training}
We train separate neural networks to perform classification and regression. For regression, we minimize the mean square error with respect to the LRMSD label (i.e., the true LRMSD). The LRMSD labels are approximately Gaussian distributed. For classification of a model as either 'acceptable' or not, we minimize the binary cross entropy. As mentioned earlier, we define an acceptable model as a model with $\mathrm{LRMSD}<10$ \AA. We weight misclassifications of acceptable models with a factor of 100 to account for the fact that most models do not fall into this category. 

For both regression and classification, we minimize the loss functions using the Adam optimizer in TensorFlow. We use Horovod \citep{Sergeev2018} to distribute training across 8 Nvidia Titan Xp GPUs. 

For both regression and classification, we repeat the stochastic optimization procedure to produce multiple machine learning (ML) models (i.e., multiple sets of optimized neural network parameters). For classification, we then select the best model based on its validation set performance with respect to the metric of \textit{success rate} (see Supporting Information). For regression, we choose the best model based on the training epoch with the smallest validation loss.

\section{Results}

\subsection{Ranking accurate models of a protein complex at the top}
\label{sec:ranking} 
The goal of a scoring function is to rank a set of structural models for a protein complex, with the most accurate models (i.e., those most similar to the unknown experimental structure) at the top. We use PAUL, the network that we trained to classify models as either acceptable or not, to remove suboptimal models from a list pre-ranked by a scoring function. Although PAUL is trained to produce a final binary output, it first computes a scalar value between 0 and 1, with larger values predictive of more accurate models. Given a list of 1000 models for a complex, pre-ranked by a scoring function, we assign a value to each model with PAUL. We then remove all models with a value below the median. \s{Figure S6 shows the effect of alternative filter thresholds on performance.} 

Here, we consider the effects of this approach on the performance of the scoring functions SOAP-PP and ZRANK. Representing two different approaches, SOAP-PP and ZRANK are representative of the most widely used methods in the field. SOAP-PP is a statistical potential based on distributions of spatial features, which was optimized with respect to models generated for DB4 \citep{Dong2013}. The features considered are interatomic distances, orientations between pairs of covalent bonds, and relative atomic surface availability. ZRANK is a physics-based scoring function that utilizes a weighted sum of seven terms accounting for electrostatic potential, van der Waals interactions, and desolvation energy at the protein-protein interface \citep{Pierce2007}.  

\begin{figure}[ht]
\begin{center}
\includegraphics[width=0.5\linewidth]{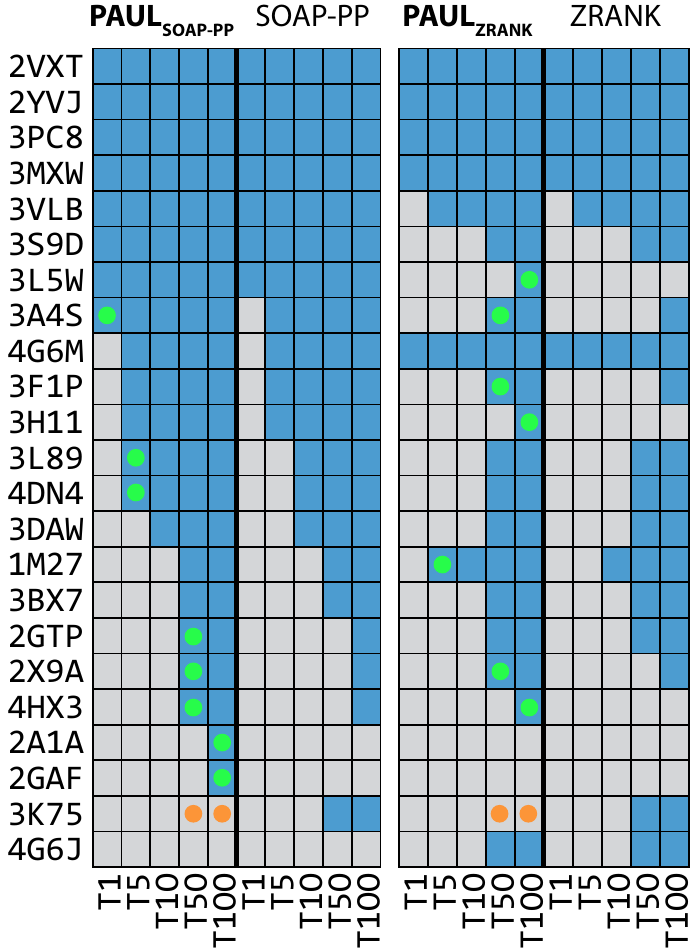}
\caption{\textbf{Ranking results per protein complex.} 
Given 1000 structural models per complex, we rank the models using the scoring functions SOAP-PP and ZRANK, with or without application of PAUL to remove suboptimal models. The figure indicates whether or not a method ranks at least one acceptable model among the first 1, 5, 10, 50, and 100 models (T1, T5, T10, T50, and T100, respectively). Blue indicates that it does, while gray indicates that it does not. \s{Green and orange dots highlight differences due to the augmentation with PAUL.} The figure shows results for 23 complexes for which the sampling software provides at least one acceptable model as part of the 1000 models and at least one of these methods ranks an acceptable model among the top 100 models.}
\label{fig:vrevenmap}
\end{center}
\vskip -0.1in
\end{figure}

Figure~\ref{fig:vrevenmap} shows the effect PAUL has on the ranking performance of SOAP-PP and ZRANK for each individual complex in the test set. Augmentation with PAUL improves performance on 8 and 7 complexes for SOAP-PP and ZRANK, respectively, while only reducing performance on a single complex.

\begin{figure}[ht]
\begin{center}
\includegraphics[width=0.75\linewidth]{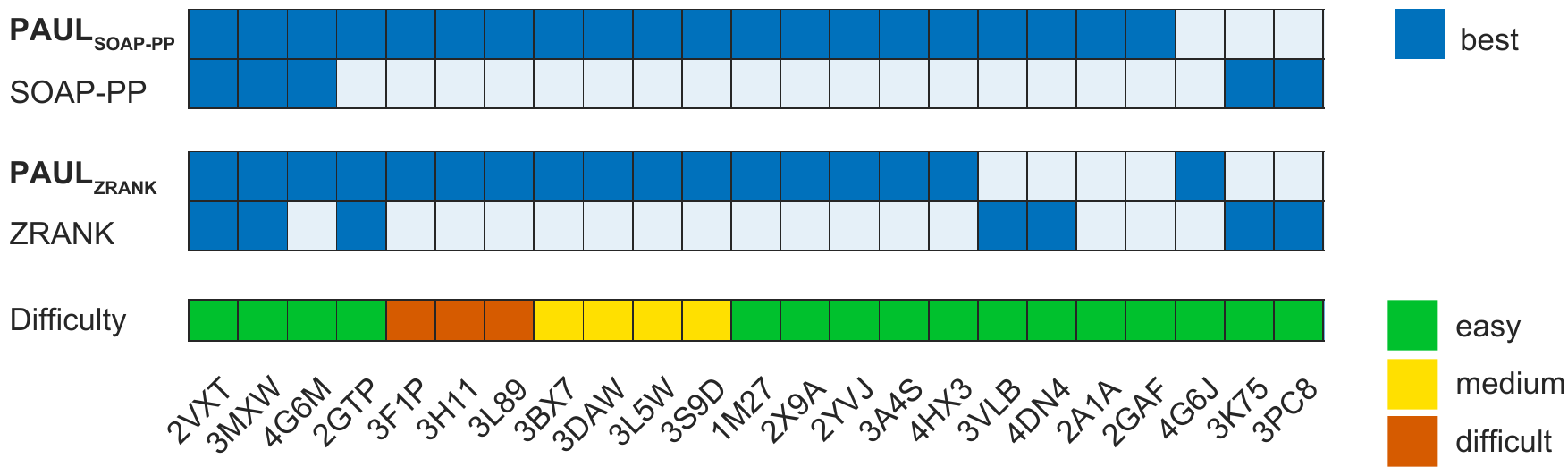}
\caption{\textbf{PAUL enriches the number of acceptable models among highly ranked models.} 
For each pair (that is, for each scoring function with and without the application of PAUL to remove suboptimal models), the best of the two ranking methods for each protein complex, as measured by the rank-weighted success metric, is shown in cyan. Both are blue when the two methods perform equally well, unless neither ranks an acceptable model among the top 100, in which case both are white. PAUL particularly enhances ranking performance for complexes in the `medium` and `difficult` difficulty categories. The categories follow the classification in \citet{Vreven2015}. 'Easy' here refers to the 'rigid-body' category.}
\label{fig:filter}
\end{center}
\vskip -0.1in
\end{figure}

Figure~\ref{fig:filter} shows the best ranking method for each complex based on the metric of \emph{rank-weighted success}.  We define \emph{rank-weighted success} $r$ as a cumulative metric:
\begin{equation}
r = \sum_{\mathrm{N} \in \left\{1,5,10,50,100\right\}} \mathrm{A(N)} \quad ,
\end{equation}
where $\mathrm{A(N)}$ is the number of acceptable models among the top-ranked $\mathrm{N}$ models.
Gains in $r$ reflect overall enrichment of acceptable models, with an emphasis on acceptable models that are ranked highly. For example, an acceptable model that is ranked 4th contributes four times as much to $r$ as an acceptable model that is ranked 51st.  

We note that the combination of PAUL with either SOAP-PP or ZRANK consistently enhances ranking performance for complexes in the 'medium' and 'difficult' categories. The difficulty categories follow the classification in \citet{Vreven2015} and are based on the deformation of the individual protein structures upon complex formation.

\s{In Supporting Information, we report results for 98 additional protein complexes from the PPI4DOCK dataset \citep{Yu2016a} (Figures S1 and S2). Augmentation with PAUL leads again to improved ranking performance for both SOAP-PP and ZRANK. This is also the case when we sample models with varying backbone conformations (Figure S3). We further show that PAUL also improves the performance of a previously designed machine-learning-based scoring function (Figure S4). Finally, we demonstrate that PAUL increases ranking performance with respect to two other metrics, success and hit rate, and we show that PAUL\textsubscript{SOAP-PP} outperforms both SOAP-PP\textsubscript{ZRANK} and ZRANK\textsubscript{SOAP-PP} (the subscript indicates the method used for ranking before filtering by the other method) (Figure S5).} 

\subsection{Predicting the absolute accuracy of a structural model} \label{sec:scatter}
We train a second rotation-equivariant neural network to predict LRMSD for a given model. Predicting LRMSD provides an estimate of how close a given model is to the correct structure (i.e., a high-resolution, experimentally determined structure). This is useful for quality assessment---that is, for assessing the likelihood that a chosen model is accurate.

\begin{figure}[ht]
\begin{center}
\includegraphics[width=0.40\linewidth]{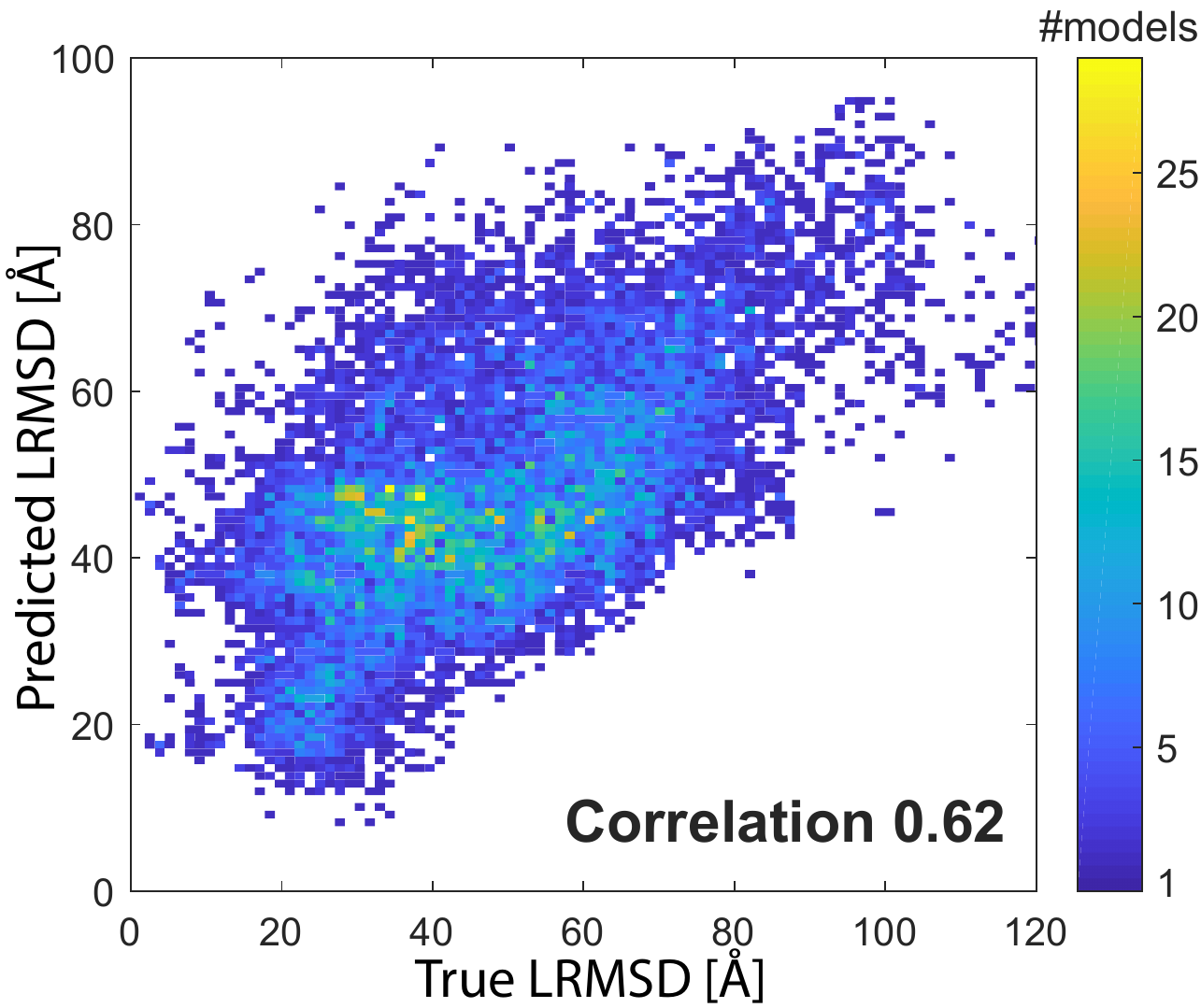}
\caption{\textbf{Correlation between predicted and actual accuracy of structural models.} 
Prediction of our trained neural network VS. true LRMSD value for diverse structural models of 50 protein-protein complexes. The color in the density scatter plot indicates the number of models per bin. Bins with no models are shown in white. Correlation indicates the Pearson correlation coefficient between predicted and true LRMSD values.}
\label{fig:scatter_scores}
\end{center}
\vskip -0.1in
\end{figure}

The density scatter plot in Figure~\ref{fig:scatter_scores} shows the predictions of our network and the true LRMSD for 300 structural models for each of 50 protein complexes. The Pearson correlation coefficient of network predictions and true LRMSD values is 0.62. \s{The correlation coefficients for ZRANK and SOAP-PP are 0.05 and 0.36, respectively; these lower correlation coefficients are not surprising, as these scoring functions are optimized to rank multiple models of the same complex rather than the prediction of absolute model quality given models for multiple complexes.}  

While relative ranking and quality assessment are separate tasks, the ability of our classifier model PAUL to remove poor structural models without removing acceptable ones may be linked to the performance of the regression model at predicting LRMSD. Figure~\ref{fig:scatter_scores} shows good correlation over a wide range of LRMSD values, including values above 50 \AA. LRMSD values in this regime generally indicate that the ligand is placed at the wrong site of the receptor. The ability to recognize such poor models likely stems from our global learning approach---learning directly from all atoms of the complex---as opposed to an approach that only locally assesses the protein-protein interface of a given structural model. 

\begin{figure}[ht]
\begin{center}
\includegraphics[width=0.9\linewidth]{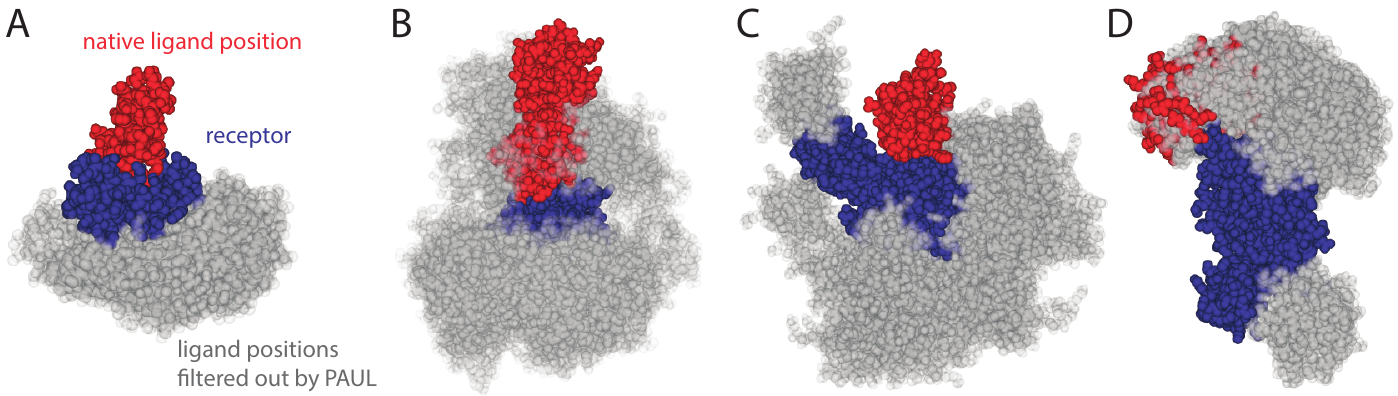}
\caption{\textbf{Effect of filtering with PAUL.}
For four different targets from our test set (corresponding to PDB entries 2X9A, 2A1A, 2GTP and 3K75 in panels A--D, respectively), we show the native protein-ligand complex (blue and red) along with the 20 highest-ranked models (as ranked by SOAP-PP) that PAUL removes (ligand in gray). PAUL improves ranking for A--C by filtering out models in which the ligand is placed distant from the actual binding site, whereas in D, PAUL mistakenly removes models in which the ligand is placed near the actual binding site.}
\label{fig:vis_filter_paul}
\end{center}
\vskip -0.1in
\end{figure}

\s{Figure~\ref{fig:vis_filter_paul} illustrates the effect of filtering with PAUL. Considering four different targets from our test set, we show the native protein-ligand complex along with the 20 highest-ranked models (as ranked by SOAP-PP) that PAUL removes. PAUL improves ranking performance for the complexes shown in panels A--C. We note that this improvement results from filtering out models in which the ligand protein is placed far away from the actual binding site. In contrast, panel D illustrates one complex for which filtering with PAUL leads to a decrease in ranking performance. In this case, PAUL mistakenly filters out all models in which the ligand is placed near the actual binding site. We make similar observations for the few targets of the extended test set in which PAUL filtering leads to a decrease in ranking performance (see Figure~S1).} 

\subsubsection{Influence of maximum rotation order and runtime measurements}
As part of our ML model selection for the experiment shown in Figure~\ref{fig:scatter_scores}, we trained 30 different networks and explored the influence of the maximum rotation order (see Section~\ref{sec:hierarchical}) on network performance. We trained 10 networks per maximum rotation order. The test losses (in units of \AA$^2$) of the best networks (selected based on the validation loss) for the maximum rotation orders 0, 1, and 2 are 290, 274, and 260, respectively. \s{The test losses for the different maximum rotation orders are significantly different (one-way ANOVA and Tukey's test with three pairwise comparisons, p<0.05).} These results are in line with the theoretical prediction that the maximum rotation order governs the amount of preserved orientation information after we apply the equivariant subsampling operation. 

For networks with equal numbers of features per point, we measured the runtimes per training epoch (mean $\pm$ std) for the maximum rotation orders of 0, 1, and 2 as $\left(290 \pm 5\right)$ s, $\left(610 \pm 10\right)$ s, and $\left(1288 \pm 7\right)$ s, respectively. We trained each network using 42,000 structural models. This is 1--2 orders of magnitude faster than voxelization-based network architectures (that only take the atoms at the protein-protein interface into account) and highlights the efficiency of our point-based approach \citep{Townshend2019}. 

\section{Discussion}
Correctly ranking candidate structures of protein complexes has remained an unsolved challenge. Difficulties include the large number of atoms involved, the inherent flexibility of the individual proteins, and the need to capture subtle interatomic interactions that are sensitive to small changes in structure. In this paper, we demonstrate how a novel machine learning approach can help to address this challenge. 

We present a deep learning architecture that can efficiently learn geometrically precise features on large systems using a rotation-equivariant neural network. This architecture is different from other machine learning approaches that have been applied to the problem of protein-protein docking  \citep{Geng2019, Wang2019, Cao2020} or to other problems in structural biology.  First, we do not provide any pre-computed physics-based energies or statistical features to the neural network. Instead, our network learns solely from the raw structural data, given by the spatial coordinates and chemical element type of each atom. Second, our architecture operates directly on point coordinates. We avoid an expensive representation of the atomic structure in terms of 3D voxels and eliminate the need for rotational data augmentation. 

Our architecture can also learn physics-based relationships involving high-order polynomials of atomic distances and multi-body interactions. 
Finally, the hierarchical design of our approach, in which we aggregate information at different levels of spatial granularity, allows us to learn end-to-end from entire protein complexes as opposed to local assessment of the protein-protein interface \citep{Geng2019, Wang2019} or learning from a more coarsely represented structure at the residue level \citep{Cao2020}. 
\nocite{Zhou2011}

We use our neural network architecture to learn a classifier, PAUL, that distinguishes acceptable from incorrect models. By removing incorrect models from a list of models ranked by a scoring function, we are able to increase ranking performance for individual complexes. PAUL particularly improves ranking performance for challenging docking targets in which the individual proteins deform substantially upon binding to one another.

Although we use PAUL to augment scoring functions in this paper, we expect a similar classifier to be useful during the model generation stage of protein docking. Specifically because PAUL rarely classifies acceptable models as incorrect, such a classifier could be used to select among candidate models before final structural refinement. 

We train a second neural network to predict LRMSD for a given structural model. The reported correlation coefficient testifies to the ability of this network to assess model quality over a wide range of LRMSD values. This ability is likely due to our global learning approach---i.e., the fact that we learn from all atoms in the complex at once, as opposed to independently assessing local regions of the protein-protein interface. We also do not require the binding affinity of a native complex to predict the accuracy of a structural model.

\s{Our empirical results support the theoretical prediction that higher-order spherical harmonics are necessary to propagate orientation information over hierarchical layers. The physics-inspired architecture remains highly efficient with respect to both data and computation, as demonstrated through the runtime measurements and the fact that PAUL was trained on fewer than 150 protein complexes. 

The point-based, rotation-equivariant architecture underlying PAUL is generally applicable to other learning tasks involving three-dimensional molecular structure and suggests benefits for further applications, in structural biology and beyond. Areas for future work include exploration of different choices for the point hierarchy used to aggregate information and of different procedures to optimize the neural network.}

\section*{Acknowledgements}
We thank Joseph Paggi, João Rodrigues, and Deniz Aydin for helpful discussions. 

\section*{Funding}
We acknowledge support from the U.S. Department of Energy, Office of Science, Office of Advanced Scientific Computing Research, Scientific Discovery through Advanced Computing (SciDAC) program, and from Intel Corporation. SE was supported by a Stanford Bio-X Bowes fellowship. RJLT was supported by the U.S. Department of Energy, Office of Science Graduate Student Research (SCGSR) program. NT was supported by the Air Force Office of Scientific Research (FA9550-16-1-0082).
\clearpage
\bibliographystyle{abbrvnat}
\bibliography{library_mod3}


\renewcommand{\thefigure}{S\arabic{figure}}
\setcounter{figure}{0}    

\clearpage
\section*{Supporting Information}
\section*{Main dataset and sampling procedure}
\label{sec:db5}
We use a subset of complexes from docking benchmark 4 (DB4)~\citep{Hwang2010} for training (142 complexes) and validation (21 complexes). We test on 50 complexes that are part of docking benchmark 5 (DB5)~\citep{Vreven2015} but not of DB4.

The training set contains 101 complexes classified as rigid, 20 complexes classified as medium, and 21 complexes classified as difficult. The difficulty categories follow the classification in \citet{Vreven2015} and are based on the deformation of the individual protein structures upon complex formation.  

We use unbound protein structures to sample protein complex models with ATTRACT \citep{Schindler2017}. All models contain hydrogens --- ATTRACT uses PDB2PQR \citep{dolinsky2004pdb2pqr} to model in missing hydrogens ---  and our neural networks are trained with 300 models per protein complex. Except when specified otherwise, all results refer to structural models sampled with fixed backbones. We report results for structural models sampled with varying backbone conformations as part of this supporting document.

The test set does not contain complexes that are homologous to complexes in the training set. A single protein complex (PDB 4DN4) in the test set contains protein constituents with homologs in the training set, but these proteins are not bound to one another in the training set. We assessed homology at the level of $30\%$ sequence identity.  
We report results for additional protein complexes from the PPI4DOCK dataset \citep{Yu2016a} in the following section. 

\section*{Results for complexes from the PPI4DOCK dataset}
\label{sec:ppi4dock}

\begin{figure}[ht]
\begin{center}
\includegraphics[width=0.6\linewidth]{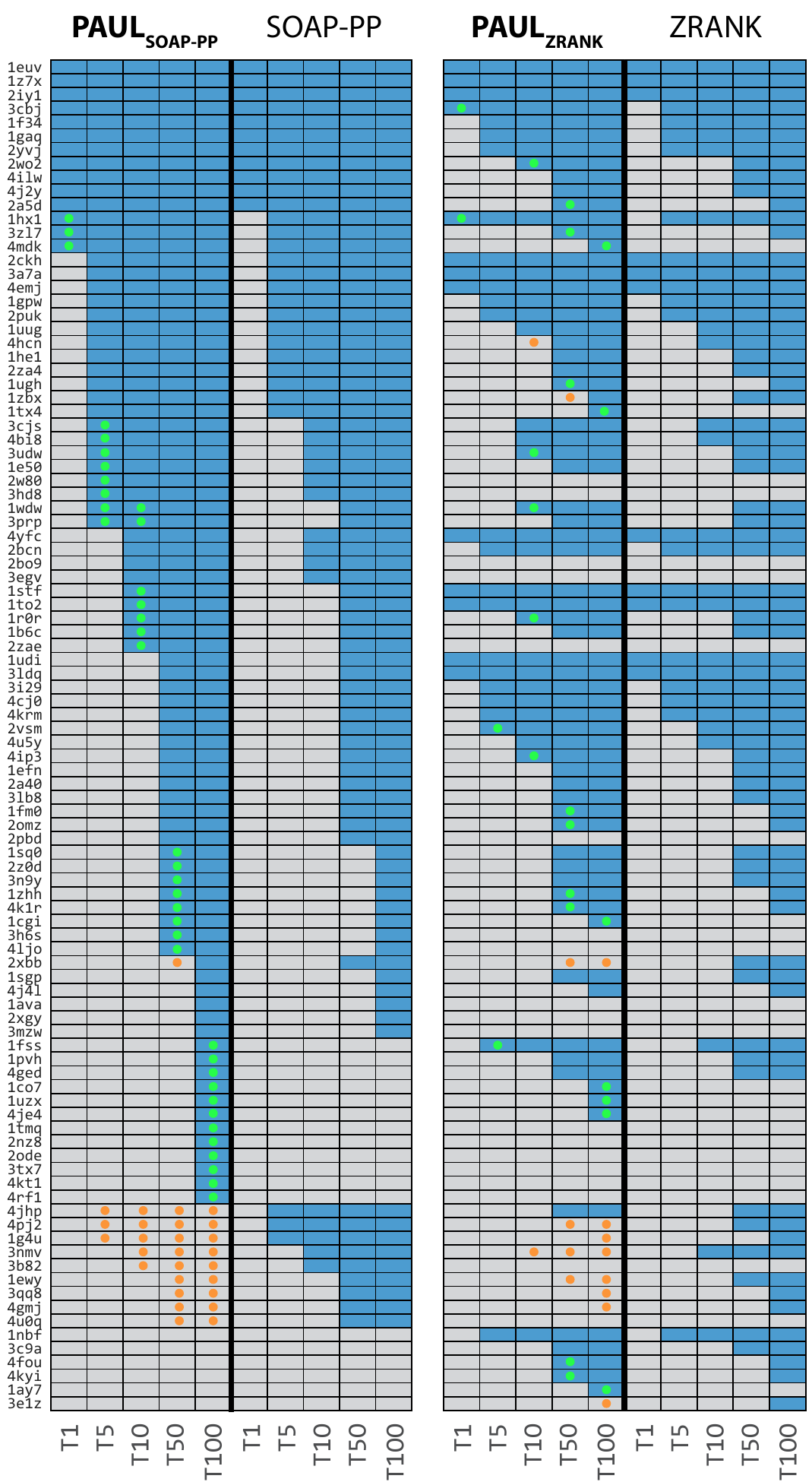} 
\caption{\textbf{Ranking results for protein complexes from PPI4DOCK.} Given 1000 structural models per complex, we rank the models using the scoring functions SOAP-PP and ZRANK, with or without application of PAUL to remove suboptimal models. The figure indicates whether or not a method ranks at least one acceptable model among the first 1, 5, 10, 50, and 100 models (T1, T5, T10, T50, and T100, respectively). Blue indicates that it does, while gray indicates that it does not. Green and orange dots highlight differences due to the augmentation with PAUL.}
\label{fig:ppi4_vreven}
\end{center}
\vskip -0.1in
\end{figure}

\begin{figure}[ht]
\begin{center}
\includegraphics[width=0.4\linewidth]{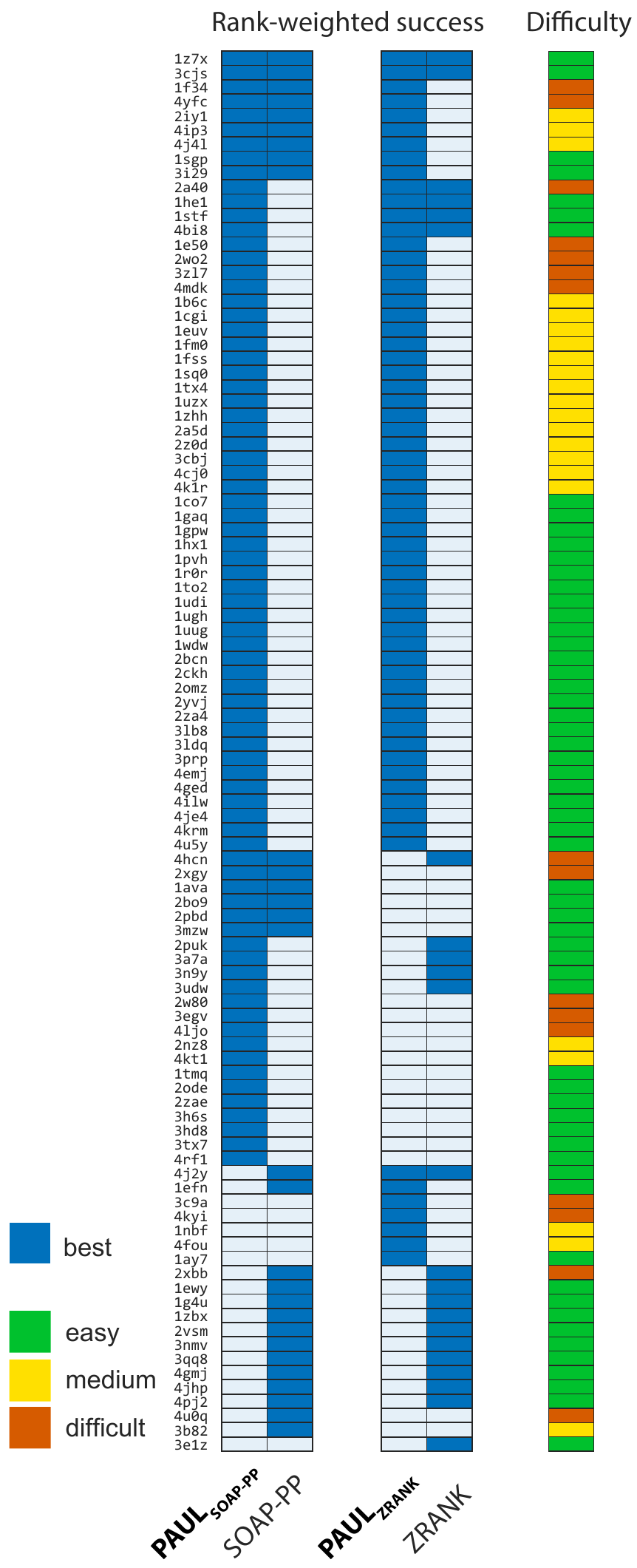} 
\caption{\textbf{Rank-weighted success for complexes from PPI4DOCK.} 
For each pair (that is, for each scoring function with and without the application of PAUL to remove suboptimal models), the best of the two ranking methods for each protein complex, as measured by the rank-weighted success metric, is shown in blue. Both are blue when the two methods perform equally well, unless neither ranks an acceptable model among the top 100, in which case both are white. The difficulty categories follow the classification in \citet{Vreven2015} and reflect the extent of protein deformation upon binding. 'Easy' here refers to the 'rigid-body' category.}
\label{fig:ppi4_rws}
\end{center}
\vskip -0.1in
\end{figure}

We report results for 98 additional protein complexes from the PPI4DOCK dataset \citep{Yu2016a} in Figures~\ref{fig:ppi4_vreven} and \ref{fig:ppi4_rws}. PPI4DOCK consists of experimentally resolved complex structures and homology models of the individual unbound proteins. 

We followed our previous protocol and sampled structural models for each target using ATTRACT \citep{Schindler2017}. We selected targets with chains A and B for which the unbound protein structures were modeled based on homologous proteins with at least 70$\%$ sequence identity. We excluded targets in which both individual proteins had more than 30$\%$ sequence identity with proteins from our training and validation set (from docking benchmark 4), and our main test set (from docking benchmark 5). 

The PPI4DOCK dataset furthermore provides a “biological” score and an “obligate” score for each complex. These scores, which are both in the range [0..1], aim to differentiate biological interactions from crystal packing contacts (“biological”) and obligate interactions from non-obligate interactions (“obligate”), respectively. We did not select complexes for sampling with a biological score below 0.5 and an obligate score above 0.5. 

\section*{Results for models sampled with flexibility}
\label{sec:ppi4dock_flex}

We sampled models with varying backbone conformations for all protein complexes in the “medium” and “difficult” category from Figure~\ref{fig:ppi4_rws} (36 complexes). The difficulty categories follow the classification in \citet{Vreven2015} and reflect protein deformation upon binding. To allow for backbone flexibility, we chose 5 normal modes for both the ligand and the receptor protein. The choice of 5 modes is the default when configuring a flexible docking protocol with ATTRACT. 

\begin{figure}[ht]
\begin{center}
\includegraphics[width=0.4\linewidth]{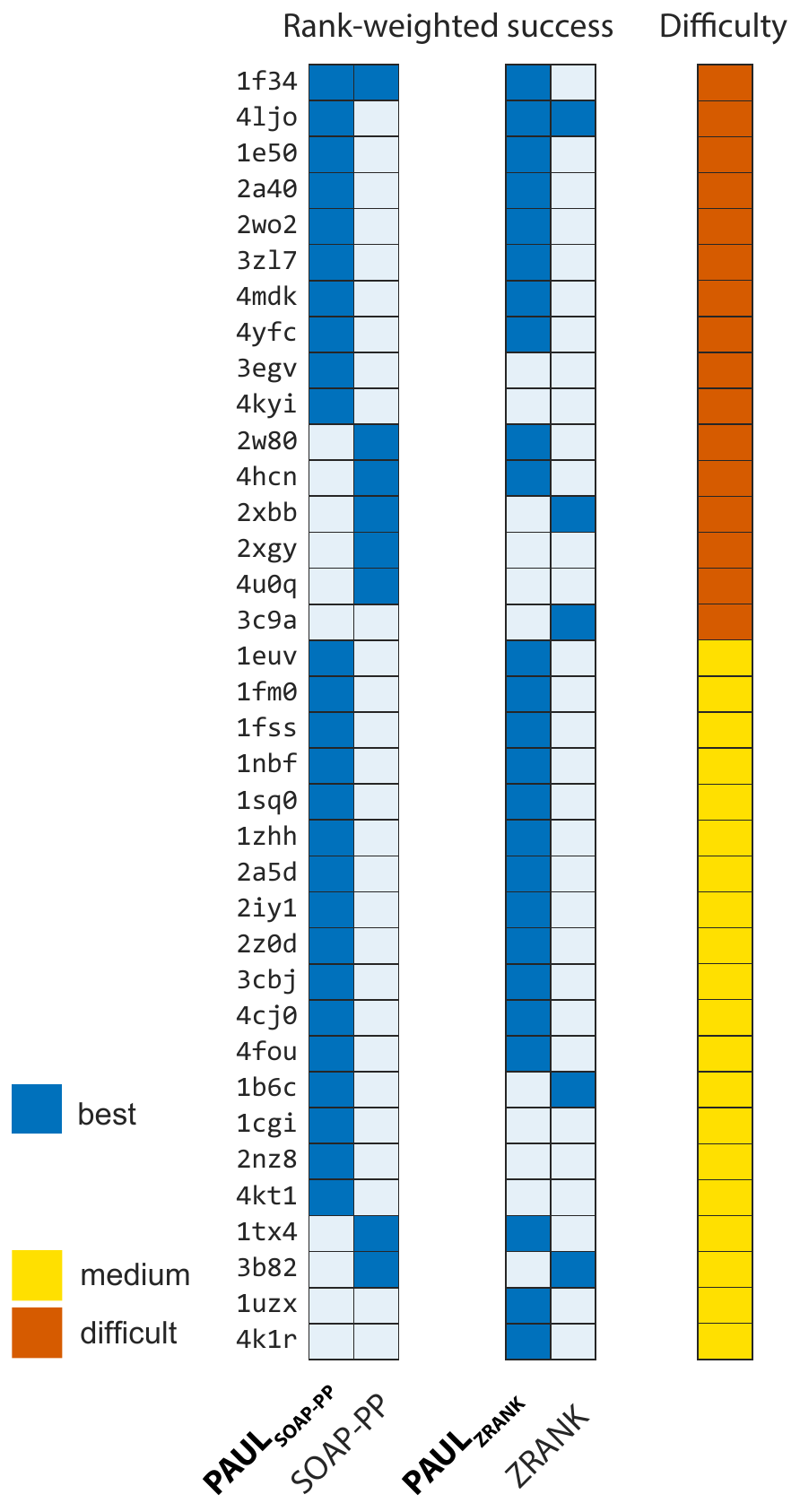}
\caption{\textbf{Rank-weighted success for complexes from PPI4DOCK sampled with flexible protein conformations.} Figure similar to Figure~\ref{fig:ppi4_rws} but for structural models sampled with flexibility (5 harmonic modes for both the receptor and the ligand protein). For each pair (that is, for each scoring function with and without the application of PAUL to remove suboptimal models), the best of the two ranking method for each protein complex, as measured by the rank-weighted success metric, is shown in blue. Both are blue when the two methods perform equally well, unless neither ranks an acceptable model among the top 100, in which case both are white. The difficulty categories follow the classification in \citet{Vreven2015} and reflect protein deformation upon binding. }
\label{fig:ppi4_flex}
\end{center}
\vskip -0.1in
\end{figure}

For these flexible models, we compare our method to SOAP-PP and ZRANK in Figure~\ref{fig:ppi4_flex}. PAUL improves upon SOAP-PP for 25 complexes, and performance decreases for 7 complexes. For ZRANK, PAUL leads to an improvement for 24 complexes, and performance decreases for 4 complexes. For 4 complexes (SOAP-PP) and 8 complexes (ZRANK), PAUL neither improves nor degrades performance.

We note that PAUL was trained with fixed-backbone models. It is possible that a version trained on flexible models would further improve prediction performance for flexible models. 

\section*{Augmenting the scoring functions DOVE and GOAP}
\label{sec:vreven_si}
In analogy to Figure~2 of the main manuscript, we use PAUL to augment the scoring functions DOVE \citep{Wang2019} and GOAP \citep{Zhou2011}. DOVE is a machine learning scoring function that uses a voxel-based convolutional neural network to locally assess the protein-protein interface of a given model. Inputs to the neural network include atomic interaction types and their energetic contributions. GOAP is a more traditional statistical scoring function similar to SOAP-PP. It is used as a reference scoring function in the DOVE publication \citep{Wang2019}.  

To augment a scoring function, we use PAUL to remove suboptimal models from a list which has been pre-ranked with the scoring function. Given a list of 1000 models for a complex, we assign a value to each model with PAUL. We then remove all models with a value below the median. 

\begin{figure}[ht]
\begin{center}
\includegraphics[width=0.4\linewidth]{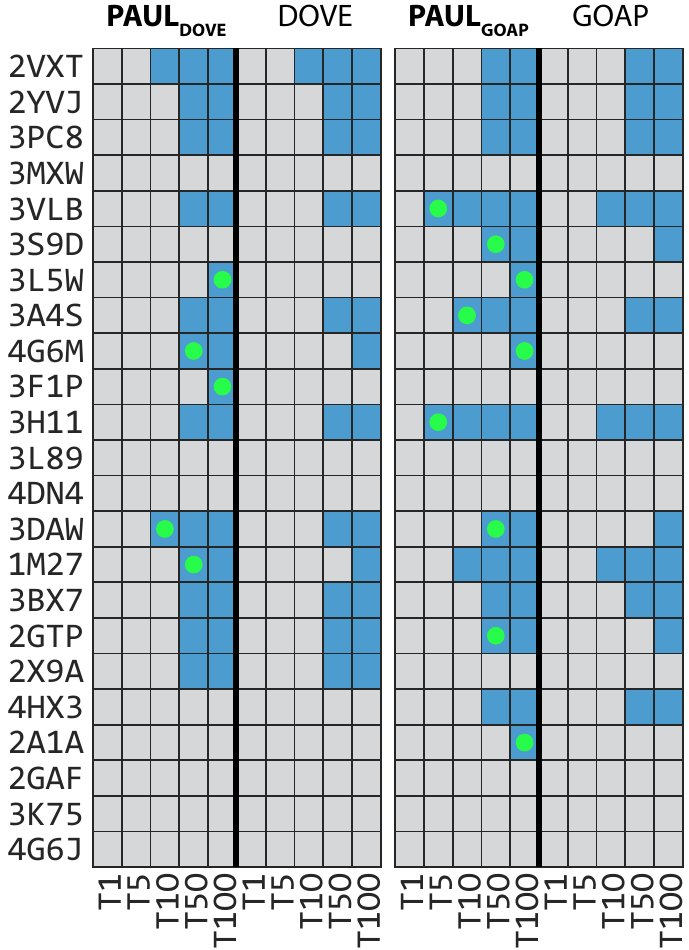} 
\caption{\textbf{Ranking results per protein complex.} Given 1000 structural models per complex, we rank the models using the scoring functions DOVE and GOAP, with or without application of PAUL to remove suboptimal models. The figure indicates whether or not a method ranks at least one acceptable model among the first 1, 5, 10, 50, and 100 models (T1, T5, T10, T50, and T100, respectively). Blue indicates that it does, while gray indicates that it does not. Green dots highlight differences due to the augmentation with PAUL. The figure shows results for the same 23 complexes as in Figure~2.}
\label{fig:si_dove}
\end{center}
\vskip -0.1in
\end{figure}

Figure~\ref{fig:si_dove} shows the effect PAUL has on the ranking performance of DOVE and GOAP for each individual complex in the test set. Augmentation with PAUL improves performance on 5 and 9 complexes for DOVE and GOAP, respectively, without reducing performance for a single complex. 

\section*{Success and hit rate}
We evaluate the effect PAUL has on ranking performance with respect to two others metrics, success and hit rate.

\subsection*{Definition of metrics}
\label{sec:def_hit_rate}
Let $\mathrm{A(N)}$ be the number of acceptable models, defined as models with $\mathrm{LRMSD}<10$ \AA, 
in a set of $\mathrm{N}$ models selected from the top of a ranked list of $\mathrm{N_{max}}$ models. \\

\noindent \textit{Success rate} $s(\mathrm{N})$ indicates whether there is at least one acceptable model among the $\mathrm{N}$ models: 
\begin{equation}
  s(\mathrm{N}) =\left\{
  \begin{array}{@{}ll@{}}
    1, &  \mathrm{A(N)} > 0 \\
    0, & \mathrm{A(N)} = 0
  \end{array}\right.  
\end{equation}

\noindent \textit{Hit rate} $h(\mathrm{N})$ describes the relative number of acceptable models $\mathrm{A(N)}$ among the top-ranked $\mathrm{N}$ models for each scoring function:
\begin{equation}
    h(\mathrm{N}) = \frac{\mathrm{A(\mathrm{N})}}{\min\left(\mathrm{N}, \mathrm{A(N_{max})}\right)} 
\end{equation}
 An ideal scoring function has the property of $h(\mathrm{N})=1$ for all $\mathrm{N}$.  

\subsection*{Different score combinations for filtering}
\begin{figure}[ht]
\begin{center}
\includegraphics[width=0.8\linewidth]{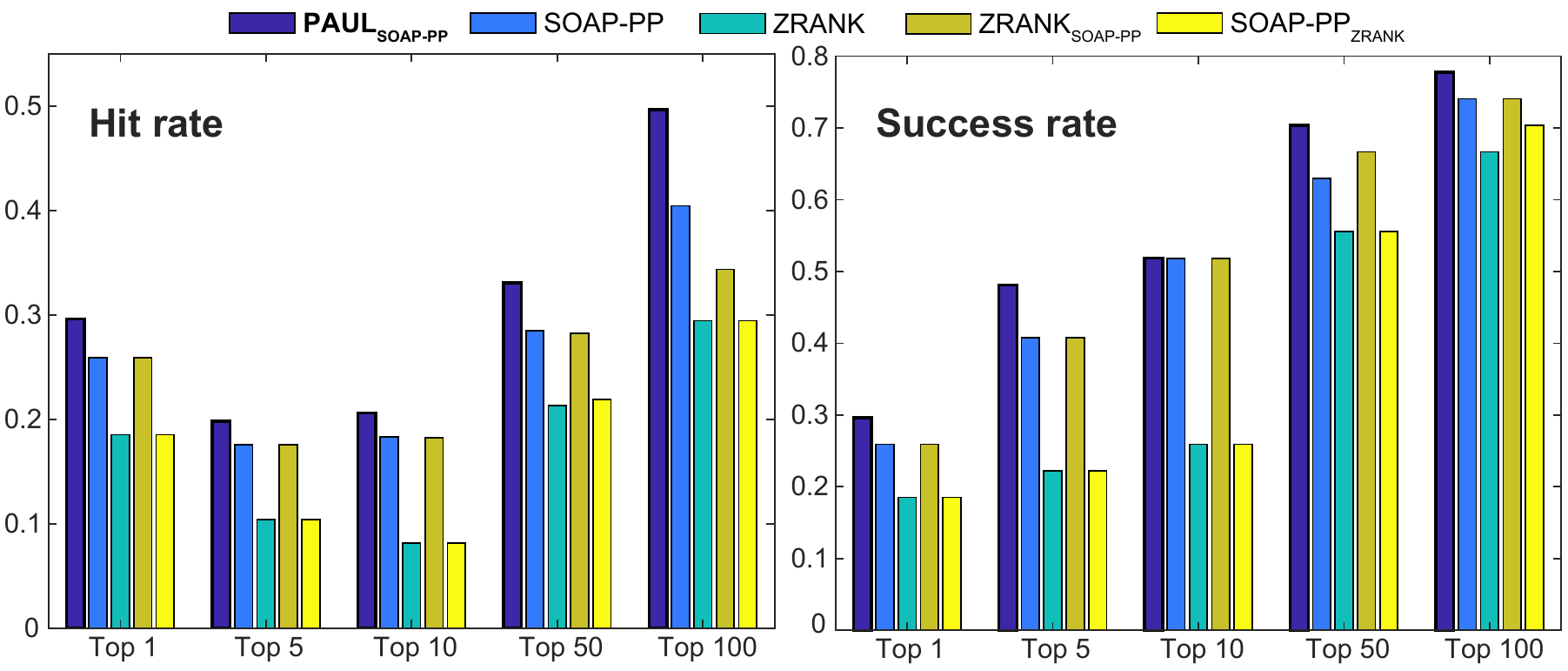} 
\caption{\textbf{Hit and success rate as a function of number of top-ranked models considered.} The values are averages for 23 complexes in the test set from docking benchmark 5 (see Figures 2 and 3).}
\label{fig:hit_rate}
\end{center}
\vskip -0.1in
\end{figure}

We use our trained network PAUL to remove suboptimal models from a list of models pre-ranked by SOAP-PP. We describe this score combination with the notation PAUL\textsubscript{SOAP-PP}. For a given list of 1000 models per complex, we remove all models for which PAUL assigns a score below the median. Figure~\ref{fig:hit_rate} shows the effect of this method for our test set. We are able to increase success and hit rate consistently (Fig.~\ref{fig:hit_rate}). The values shown are computed as averages over 23 complexes in the main test set (from DB5) with at least one acceptable model among the first 1000 models generated by ATTRACT. 
We also demonstrate that other combinations of filtering, namely SOAP-PP\textsubscript{ZRANK} and ZRANK\textsubscript{SOAP-PP}, do not result in improved success and hit rates. 

\subsection*{Varying the filter cutoff}

\begin{figure}[ht]
\begin{center}
\includegraphics[width=0.8\linewidth]{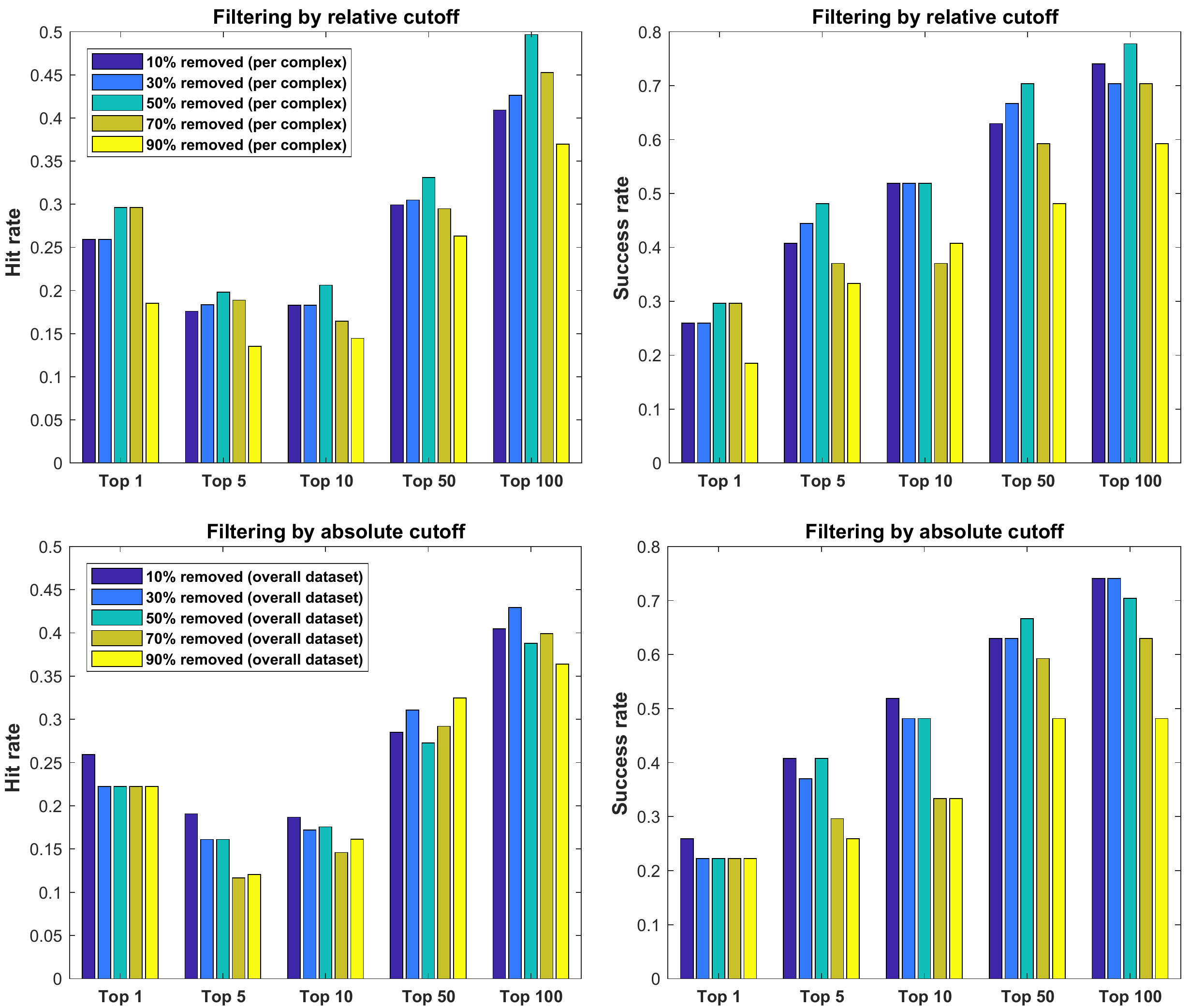} 
\caption{\textbf{Effect of filter cutoff on hit and success rate.} In the case of relative filtering (upper panels), we choose the filter cutoff based on removing a specified fraction of models per complex. In the case of absolute filtering (lower panels), we choose this filter cutoff based on the overall distribution of PAUL outputs for all models in the test set. The values are averages for 23 complexes in the test set from docking benchmark 5 (see Figures 2 and 3).}
\label{fig:vary_filter}
\end{center}
\vskip -0.1in
\end{figure}

We show the effect of varying the filter cutoff that we use to remove models in Figure~\ref{fig:vary_filter}. 

The two upper panels show results for different levels of a \emph{relative} cutoff. The default is that we remove all models for a given complex for which PAUL assigns a score below the median. Hit and success rate increase as the number of removed models increases from 10$\%$. Both metrics decrease as we remove more than 50$\%$ of models per complex. We note that even with 90$\%$ of all models per complex removed, the observed drop-off is relatively mild.

The two lower panels show results for different levels of an \emph{absolute} cutoff based on the PAUL score. Unlike the \emph{relative} cutoff, the \emph{absolute} cutoff does not depend on the distribution of PAUL outputs for each individual complex. We note that the relative filtering approach does better overall. This is in line with our expectations, as we assess ranking performance individually for each complex \textit{i.e}., we only consider models for one complex at a time.

\section*{Nearest-neighbor convolutions}
\label{sec:knn}
\begin{figure}[ht]
\begin{center}
\includegraphics[width=0.4\linewidth]{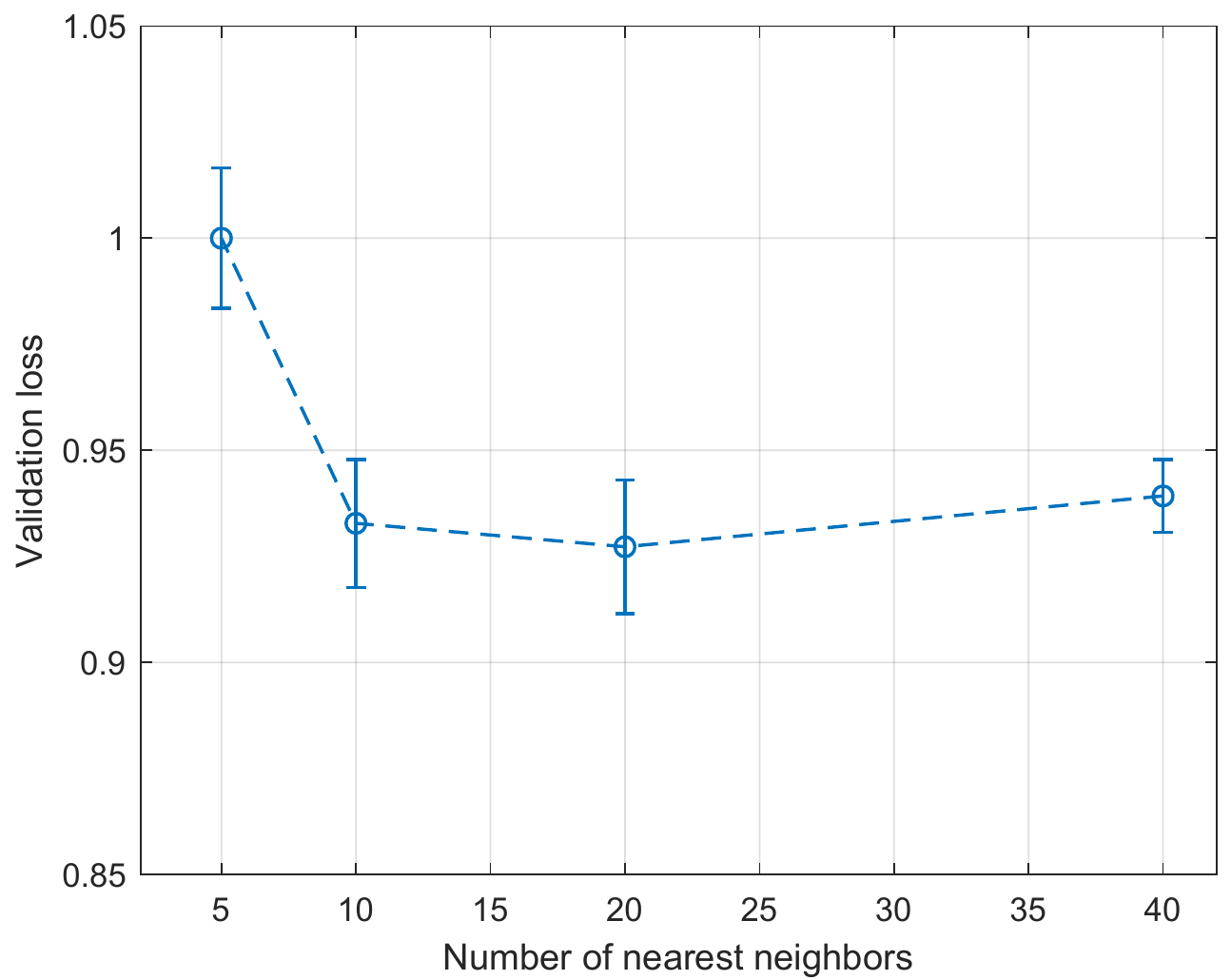}
\caption{\textbf{Varying the size of nearest-neighbor convolutions.} We trained a set of models for which we varied the number of nearest neighbors (K) that we considered for convolutions. We trained 5 models for each chosen K value, and we show mean and standard error of the mean for the validation loss of these 5 models. Apart from K, all parameters and settings are identical to our training for PAUL. Values are normalized with respect to the mean at K=5.}
\label{fig:knn}
\end{center}
\vskip -0.1in
\end{figure}

We perform nearest-neighbor convolutions with K=40 neighbors at all levels of point hierarchy and for both PAUL and the regression network trained to predict LRMSD. We trained a set of models with varying K values (but otherwise identical settings to our training for PAUL) and show the results in Figure~\ref{fig:knn}. We trained 5 models for each chosen K value. The results seem to indicate that the specific choice of K is not critical. The reasoning behind K=40 was our desire to capture all potentially relevant information through relatively large, overlapping convolution regions, while also considering the increased memory requirements that come with increases in K. In general, K could be included in the general hyperparameter optimization. 

\end{document}